\documentclass[ reprint, amsmath,amssymb,aps,superscriptaddress]{revtex4-1}
\usepackage{graphicx}% Include figure files
\usepackage{dcolumn}% Align table columns on decimal point
\usepackage{bm}% bold math
\usepackage{color}
\begin{document}

\title{The Coulomb potential $V(r)=1/r$ and other radial problems on the Bethe lattice}

\author{Olga Petrova}
\affiliation{Max Planck Institute for the Physics of Complex Systems, 01187 Dresden, Germany}
\author{Roderich Moessner}
\affiliation{Max Planck Institute for the Physics of Complex Systems, 01187 Dresden, Germany}

\begin{abstract}

We study the problem of a particle hopping on the Bethe lattice in the presence of a Coulomb potential. We obtain an exact solution to the particle's Green's function along with the full energy spectrum. In addition, we present a mapping of a generalized radial potential problem defined on the Bethe lattice to an infinite number of one dimensional problems that are easily accessible numerically. The latter method is particularly useful when the problem admits no analytical solution.

\end{abstract}
\maketitle

\section{Introduction}

In many body physics, exactly solvable models are few and far between. In their absence, a common strategy to pursue is the use of mean field approximations \cite{Weiss1907}, where the interactions between a finite number of the system's constituents and the rest are modeled through an effective field approximating the effects of the latter on the former. One of such approaches bears the name of its inventor, Hans Bethe \cite{Bethe1935}, and turns out to be exact on cycle-free graphs \cite{Kurata1953}. This coined the term ``Bethe lattice'', which, despite its seeming unphysicalness, has since been successfully used to describe a plethora of physical phenomena, including excitations in antiferromagnets \cite{Brinkman}, Anderson localization \cite{0022-3719-6-10-009}, percolation \cite{percolation}, and hopping of ions in ice \cite{Chen} to name a few. In particular, the Coulomb potential problem on the Bethe lattice, first appearing in  \cite{Gallinar} and addressed in this work, recently surfaced in a study of quantum spin ice by the authors \cite{diluted}. 

In this paper we treat the problem of a single particle hopping on the Bethe lattice in the presence of a radial potential:
\begin{equation}
H=T+V(n)
\label{eq:Hgen}
\end{equation}
where $T$ is $-t$ times the adjacency matrix, and the potential $V(n)$ is a function of the Bethe lattice generation $n$ only -- i.e., a \emph{radial} potential. The two central results of this work are: (a) a mapping of the general radial $V(n)$ problem to a family of one dimensional chains, and (b) the exact solution for the attractive Coulomb potential, $V(n)=\frac{C}{n}$ where $C<0$. The latter comes in the form of a closed form expression for the lattice Green's function. With that, one can obtain the energy levels of the model, and the local density of states as a function of $n$.

The mapping of the Bethe lattice model (\ref{eq:Hgen}) to a family of $1d$ problems, is particularly useful when the exact solution to the problem cannot be obtained. Normally, numerical treatments of Bethe lattice models are of limited use owing to the large fraction (no less than half the total) of vertices at the edges. Mapping to $1d$ chains lets us circumvent this obstacle.

The paper is structured as follows. In Section \ref{sec:tightbinding} we discuss some features common to all models where a single particle is hopping on the Bethe lattice, or a finite Cayley tree, in the presence of a radial potential: the symmetries of the Hamiltonian (\ref{eq:Hgen}) (\ref{sec:symmetries}), the mapping of different symmetry sectors to a family of $1d$ problems (\ref{sec:1Dmapping}), and the continued fraction technique to solving Eq. (\ref{eq:Hgen}) perturbatively (\ref{sec:greens}). In the technical heart of the paper, Section \ref{sec:coulomb}, we present the exact solution to the Coulomb potential problem on the Bethe lattice, obtained by carrying out the aforementioned perturbative calculation to infinite order. We summarize our findings and discuss possible applications in Section \ref{sec:conclusion}.   

\section{Tight-binding radial Hamiltonians on the Bethe lattice}
\label{sec:tightbinding}

In this Section we discuss some general properties of radial tight-binding models (\ref{eq:Hgen}) defined on the Bethe lattice, a rooted infinite cycle-free graph with coordination number $z$ and connectivity $K$, defined as $K=z-1$. The root vertex is labeled by 1. To simplify further discussion, we reduce the coordination number at 1 to $z-1$, as shown in Fig.~\ref{fig:lattice}(a), such that every vertex of the Bethe lattice generation $n$ is connected to $K$ vertices at generation $(n+1)$. Generations $n=1$ and $n=2$ correspond to the root and its nearest neighbors respectively. Due to the absence of closed cycles in the Bethe lattice, there is a unique path connecting any two vertices.
The distance from the root to a given node is therefore given by the node's generation $n-1$.

Much of the discussion presented here applies to finite Cayley trees as well: the symmetry analysis (\ref{sec:symmetries}) carries over directly, whereas the $1d$ chains and the continued fractions that one gets from the mapping (\ref{sec:1Dmapping}) and the expansion of the Green's function in powers of $t$ (\ref{sec:greens}) respectively, are finite, rather than infinite, for the Cayley tree problems. One particularly simple result that follows is that a Cayley tree with $M$ generations has $K^{M-2}$ degenerate edge modes with energy $V(M)$.

\subsection{Symmetries of the model}
\label{sec:symmetries}

Here and in Section \ref{sec:1Dmapping}, we treat the $z=3$ case for concreteness unless stated otherwise. We refer to two vertices at generation $(r+1)$ as \emph{siblings} if they are connected to the same \emph{parent} vertex at generation $r$. %A \emph{branch} is a set of all descendants of a given vertex.

Exchanging left and right siblings at a given generation leaves the Hamiltonian (\ref{eq:Hgen}) invariant \cite{Chen}. These operations commute, which allows us to associate a separate quantum number, denoting the parity under such exchanges, with each generation $n>1$. Each such symmetry sector can be mapped onto a $1d$ half-line, whose origin is offset by the number equal to the highest Bethe lattice generation with an odd quantum number for a given sector. 

\begin{figure}
\begin{center}
\includegraphics[width=0.9\linewidth]{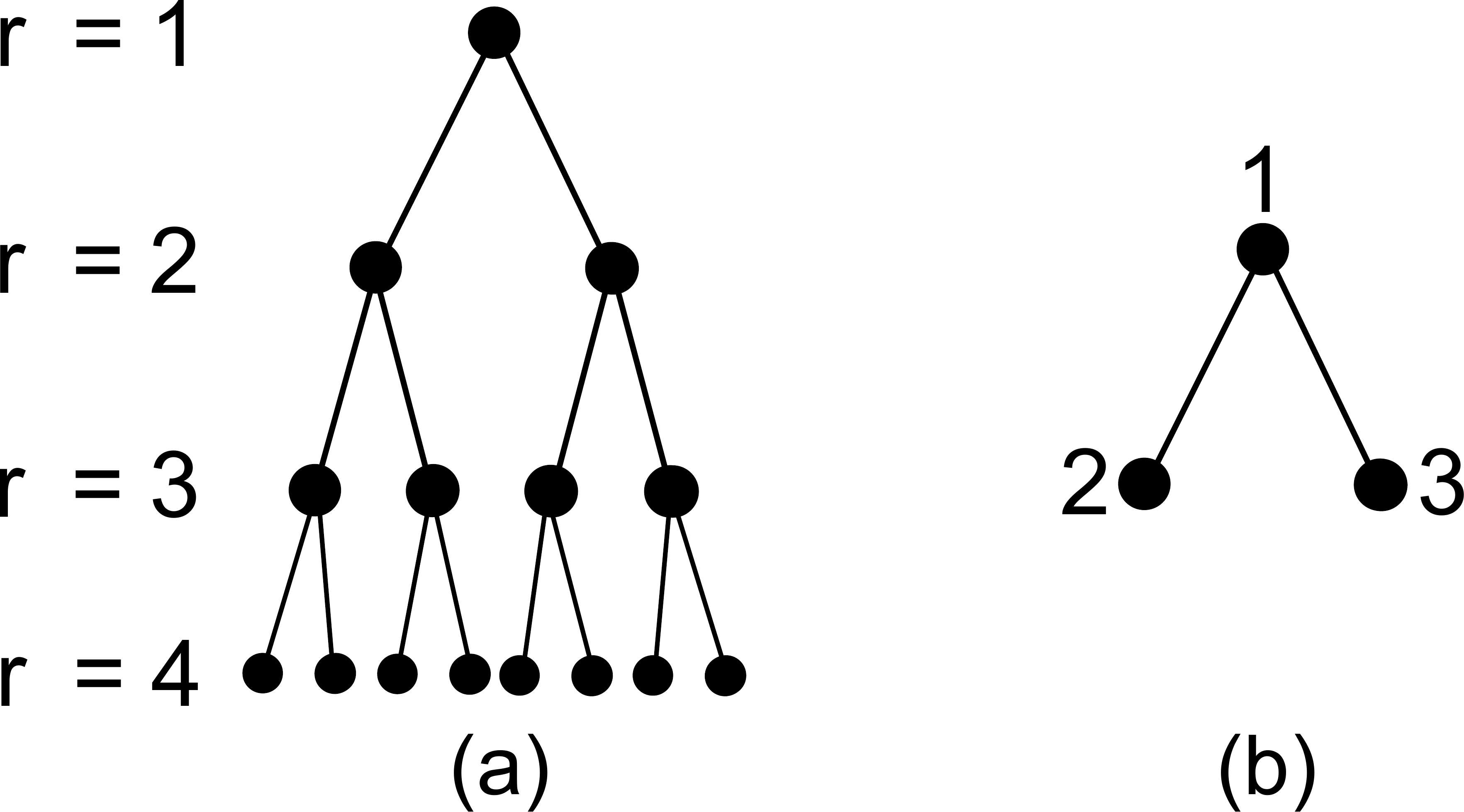}
\end{center}
\caption{Left: the first four generations of the Bethe lattice with $z=3$. The origin ($r=1$) has coordination number $z-1$. Right: a three site fragment of the Bethe lattice.}
\label{fig:lattice}
\end{figure}

To understand the physical meaning of the mapping to follow, consider three vertices depicted in Fig.~\ref{fig:lattice}(b): parent site 1 at generation $r$, and siblings 2 and 3 at generation $r+1$ that are connected to it. Let us construct two states: a symmetric and an antisymmetric combination of the particle being at sites 2 and 3, $|\Psi_S\rangle$ and $|\Psi_A\rangle$ respectively. When the hopping term $T$ in Hamiltonian (\ref{eq:Hgen}) acts on $|\Psi_S\rangle$, the contributions from sites 2 and 3 add up enabling the particle to hop back to the parent site 1 with amplitude $2t$. By contrast, $T$ acting on $|\Psi_A\rangle$ results in back-hopping canceling out. From the extension of this argument to the entire generation of vertices rather than a single sibling pair, it follows that a particle starting out at a state that is odd at the $k^\mathrm{th}$ generation cannot hop to generations $r<k$. Therefore, the particle's wavefunction has zero amplitude at all generations $r<k$: in particular, only the all even states have a nonzero amplitude at the origin. This property allows us to make a statement about the degeneracies of the energy levels: if the sector's highest generation with an odd quantum number is $n$, then each of the lower generations $r>2$ contributes a factor of 2 to the total degeneracy of that sector's energy levels.

\subsection{Mapping to infinite half-lines}
\label{sec:1Dmapping}

We now describe how different Hilbert space sectors, labeled by the even and odd quantum numbers discussed above, can be mapped onto one-dimensional chains. The connection between the spectra of a Hamiltonian defined on a half-line and a related problem on the Bethe lattice was known previously \cite{ASW}. The technique we present can be used to calculate the full spectrum of the Bethe lattice problem from $1d$ models that have the advantage of being trivial to simulate numerically. Moreover, one can also use it to obtain the eigenstates of the original problem via exact diagonalization, or their probabilities by the analytic method described in Section \ref{sec:greens}.

The mapping can be understood in the following way. We start in the $|i\rangle$ basis, where $i$ identifies individual  nodes of the Bethe lattice. Having labeled the vertices in Fig.~\ref{fig:lattice}(a) from left to right at each row, starting with $1$ for the root, we end up with a Hamiltonian matrix of the following form:
\begin{equation}
H = \begin{bmatrix}
       V(1) & -t & -t & 0  & ...        \\
       -t & V(2) & 0 & -t & ...\\
       -t & 0 & V(2) & 0 & ... \\
       0 & -t & 0 & V(3) & ... \\
 		... & ... & ... & ... & ...
     \end{bmatrix}.
     \label{eq:mappingH1}
\end{equation}
The matrix above can be brought to a block diagonal form via a unitary transformation into the basis of states defined by their even/odd quantum numbers:
\[
H = \begin{bmatrix}
       V(1) & -\sqrt{2}t &0 & 0  &0&0& ...        \\
       -\sqrt{2}t & V(2) & -\sqrt{2}t & 0&0 &0& ...\\
       0 & -\sqrt{2}t & V(3) &  -\sqrt{2}t &0&0& ... \\
  		 ...&...&...&...&...&...&... \\
	0 & 0 & 0 & 0 &V(2)&-\sqrt{2}t&... \\
	0 & 0 & 0 & 0 &-\sqrt{2}t&V(3)&... \\
	... & ... & ... & ... &...&...&...
     \end{bmatrix}
\]
where the upper left block corresponds to the all-even sector of the Hilbert space, and the next one to the states which are odd at the second generation and thus have zero amplitude at site 1. The problem of a single particle hopping on the Bethe lattice in the presence of a radial potential $V(n)$ has thus been decomposed into a series of one-dimensional problems with hopping strength multiplied by a factor of $\sqrt{2}$ compared to the original, such that in each $k^\mathrm{th}$ infinite half-line $V(n)$ is offset by $k$. The degeneracies discussed in the previous Section are reflected in the fact that the block diagonal form of the Hamiltonian (\ref{eq:mappingH1}) contains multiple identical $1d$ problems starting with the third generation, and their number is equal to the degeneracy of the respective sector.

The energy levels of the Bethe lattice problem with hopping amplitude $-t$ and radial potential $V(n)$ have a trivial relation to those of the related $1d$ chains: namely, Bethe lattice energies from the sector with highest odd generation at $k$ are given by the energies of the $1d$ chain with hopping $-\sqrt{2}t$ and potential $V(n+k-1)$. In that sector, the probability amplitude for finding the particle in a state with energy $E_n$ at Bethe lattice \emph{generation} $r$ is given by the $(r-k+1)^\mathrm{th}$ component of the eigenvector with eigenvalue $E_n$ of the corresponding $1d$ problem.

Our next step depends on whether the form of $V(n)$ in Eq.~(\ref{eq:Hgen}) allows for an exact solution. In the case that it does not, exact diagonalization will readily provide us with both the energy levels and the wavefunctions for each one-dimensional chain, which can then be used to derive the corresponding observables for the original Bethe lattice problem. The advantage of employing the mapping technique lies in the fact that for a one dimensional chain, boundary effects quickly become negligible as we increase the system size. On the other hand, Cayley trees are notoriously difficult to deal with numerically due to the large fraction of nodes at the systems' edges.

On the analytic front, there are various ways to approach one-dimensional hopping problems on the lattice. The one that we discuss in the next Section can be applied to the original Bethe lattice problem directly, in addition to the special $z=2$ case of one-dimensional chains. When the technique that we are about to discuss results in an exact solution for the lattice Green's function, there is no benefit in first decomposing the Bethe lattice problem to a series of $1d$ ones, so the generalized coordination number $z$ and connectivity $z=K-1$  will be used in the rest of the paper.

Before we turn to the discussion of the Green's function calculation, we note that both the symmetry analysis and the mapping to $1d$ have a straightforward generalization to Bethe lattices with larger coordination numbers $z>3$ \cite{Chen}. In this case, each generation's quantum number takes one of $K$ values. Going back to our three site example in Section \ref{sec:symmetries}, consider connecting site 1 in the Bethe lattice fragment in Fig.~\ref{fig:lattice}(b) to $K$, rather than 2, descendants, labeled from 2 to $K+1$. The argument remains almost identical, except for there now being $K-1$ states that do not give rise to back-hopping: $|\Psi_A^j\rangle=(|2\rangle+e^{1\times(i2\pi j/K)}|3\rangle+...+e^{(K-1)\times(i2\pi j/K)}|K+1\rangle)/\sqrt{K}$, where $j$ is an index going from $1$ to $K-1$.

\subsection{The continued fraction method for calculating the lattice Green's function}
\label{sec:greens}

The cycle-free nature of the Bethe lattice makes it well-suited for recursive approaches \cite{0022-3719-6-10-009,Brinkman}. In order to calculate the Green's function we separate the Hamiltonian (\ref{eq:Hgen}) into two parts
\begin{equation}
H_0 = V(n) \qquad \mathrm{and} \qquad H'=T
\end{equation}
and generate a perturbation series in $t$ using the Dyson equation
\begin{equation}
\mathcal{G}(\omega)=\mathcal{G}_0(\omega)+\mathcal{G}_0(\omega)\Sigma(\omega)\mathcal{G}(\omega)
\label{eq:Dyson}
\end{equation}
where $\mathcal{G}_0(\omega)$ is the unperturbed Green's function
\[
\mathcal{G}_0(\omega)=\cfrac{1}{\omega-H_0},
\]
and $\Sigma(\omega)$ is the self-energy of the particle. If we consider a diagonal element $G_i(\omega)$ of the Green's function operator, its self-energy is given by a sum of terms associated with all paths on the Bethe lattice going away from node $i$ and back to it. For each hop one such term gains a factor of $-t$ and is divided by $\omega-H_0$ evaluated at the ``arrival'' node (excluding the starting vertex $i$). The lack of closed cycles on the lattice makes all paths that we need to count self-retracing, allowing for a straightforward way to sum them up. We can write $G_i(\omega)$ as a power series expansion in $t$, or, equivalently, as a continued fraction. For instance, for the root vertex $1$ the diagonal element of the lattice Green's function is given by 
\begin{equation}
G_1(\omega)=\cfrac{1}{\omega-V(1)-\cfrac{Kt^2}{\omega-V(2)-\cfrac{Kt^2}{\omega-V(3)-...}}}.
\label{eq:G1}
\end{equation}
The self energies for other nodes of the Bethe lattice involve, in addition to ``forward'' hops to higher generations, paths that go through the root. Their Green's functions $G_{i\in n}(\omega)$ (where $i$ is a node at the $n$th generation) can be defined in terms of a finite number of $G_{k}^{F}(\omega)$, infinite continued fractions involving only the forward hops starting from a node at the $k^\mathrm{th}$ generation:
\begin{widetext}
\begin{equation}
G_{i\in n}(\omega)=\cfrac{1}{\omega-V(n)-Kt^2G^F_{n+1}(\omega)-\cfrac{t^2}{\left[G_{j\in n-1}(\omega)\right]^{-1}+t^2G^F_n(\omega)}}
\label{eq:Gfull}
\end{equation}
\end{widetext}
where
\begin{equation}
G_{k}^{F}(\omega)=\cfrac{1}{\omega-V(k)-\cfrac{Kt^2}{\omega-V(k+1)-...}}.%\cfrac{Kt^2}{\omega-V(k+2)-...}}}.
\label{eq:Gfor}
\end{equation}

Provided we can obtain a closed form expression for Eq.~(\ref{eq:Gfor}), the diagonal elements $G_i(\omega)$ of the full Green's function give us the system's energy levels (poles of $G_i(\omega)$ occur at the energies $\omega$ whose corresponding eigenstates are visible at the lattice node $i$) and the local densities of states (the local DOS at node $i$ is proportional to Im$\left[G_i(\omega)\right]$). The probability of the bound state with an energy $\omega_n$ at node $i$ is given by the residue of $G_i(\omega)$ evaluated at $\omega_n$. A continuous imaginary part of $G_i(\omega)$ points to the existence of a continuum energy band.

\subsection{Constant potential on a sublattice}

Given the particle's Green's function, we have the full knowledge of its energy spectrum, but not of its eigenstates. Although a direct way to obtain the exact eigenstates for a general radial $V(n)$ is often unavailable, there are specific cases when a large fraction of them can be deduced. One such case is that of a potential that takes a constant value on at least one of the two sublattices (corresponding to alternating generations of the Bethe lattice), e.g., $V_A$ on sublattice A. The bipartite nature of the lattice allows for a straightforward way of constructing eigenstates with the energy $V_{A}$: amplitudes on the A nodes are assigned in such a way that hopping to the B nodes cancels out. For instance, for a non-normalized eigenstate corresponding to eigenvalue $V_A$, start with amplitude $+1$ at the root (taking it to belong to sublattice A), and assign zeros on all even (B) generations and $\left(-1/K\right)^{\frac{n-1}{2}}$ on all odd (A) ones. Alternatively, we can choose a sibling set at any sublattice A generation $k$, assign amplitudes to the siblings such that hopping back to the parent node cancels out,
\[
\sum_{i=1}^Ks_i=0,
\]
and construct the rest of the state by assigning amplitude $s_i\times\left(-1/K\right)^{\frac{n-k}{2}}$ to all the A nodes in the $i^\mathrm{th}$ sibling branch, and zeros elsewhere. Such bipartite eigenstates with energy $V_A$ constitute fraction $1/2K$ of the total number of eigenstates.

\section{The Coulomb potential problem}
\label{sec:coulomb}

We consider a single particle hopping on the Bethe lattice in the presence of an attractive Coulomb potential:
\begin{equation}
H=T+C/n
\label{eq:Hcoul}
\end{equation}
where $T$ is the hopping matrix whose non zero elements are equal to $-t$, $n$ is the generation of the Bethe lattice ($n=1$ at the root), and $C<0$. 

\subsection{The exact solution}
\label{sec:exact}

What makes the Coulomb problem exactly solvable is the fact that the infinite continued fractions of the form (\ref{eq:Gfor}) with $V(n)=C/n$ can be written in closed form, with the use of special functions \cite{Ramanujan}:
\begin{widetext}
\begin{equation}
G^{F}_k(\omega)=\cfrac{2k/\omega}{\sqrt{1+x^2}+1}
\cfrac{1}{k-\cfrac{C/\omega}{\sqrt{1+x^2}}}
\cfrac{F^2_1\left(1-\cfrac{C/\omega}{\sqrt{1+x^2}},k+1,k+1-\cfrac{C/\omega}{\sqrt{1+x^2}},\cfrac{1-\sqrt{1+x^2}}{1+\sqrt{1+x^2}}\right)
}{F^2_1\left(1-\cfrac{C/\omega}{\sqrt{1+x^2}},k,k-\cfrac{C/\omega}{\sqrt{1+x^2}},\cfrac{1-\sqrt{1+x^2}}{1+\sqrt{1+x^2}}\right)}
\label{eq:Gexact}
\end{equation}
\end{widetext}
where $x^2=-\cfrac{4Kt^2}{\omega^2}$ and $F^2_1(a,b,c,z)$ is Gauss hypergeometric function. We can thus obtain the diagonal elements of the Green's function $\mathcal{G}(\omega)$ for our problem exactly, and use them to calculate the energy levels and the local densities of states, which we do next.

\subsection{The all-even bound states}
\label{sec:alleven}

We already know from Section \ref{sec:symmetries} that only the all-even states are allowed at the root. Therefore, all the energy levels of the all-even sector are given by the poles of $G_1(\omega)=G^{F}_1(\omega)$. Obtained from Eq.~(\ref{eq:Gexact}), it is proportional to a ratio of two hypergeomeric functions. In the $k=1$ case, the hypergeometric function in the denominator has no zeros. The ground state energy corresponds to the pole that $G_1(\omega)$ has when $1-\cfrac{C/\omega}{\sqrt{1+x^2}}$ in the denominator is equal to zero, whereas the
excited energy levels are given by the poles of the hypergeometric
function in the numerator:
\[
G_1(\omega)\propto
F^2_1\left(1-\cfrac{C/\omega}{\sqrt{1+x^2}},2,2-\cfrac{C/\omega}{\sqrt{1+x^2}},\cfrac{1-\sqrt{1+x^2}}{1+\sqrt{1+x^2}}\right).
\]
To find them, we make use of the hypergeometric series representation of $F^2_1(a,b,c,z)$:
\[
F^2_1(a,b,c,z)=\sum^{\infty}_{j=0}\cfrac{(a)_j(b)_j}{(c)_j}\cfrac{z^2}{j!}
\]
where $(x)_j$ is the Pochhammer symbol $(x)_j=x(x+1)...(x+j-1)$. Evidently, $F^2_1(a,b,c,z)$ has a pole whenever $c+j=0$, where $j$ is an integer from $0$ to $\infty$. Together with the pole when $1-\cfrac{C/\omega}{\sqrt{1+x^2}}=0$, this leads to the following energy levels:
\begin{equation}
\omega_n=-\sqrt{\cfrac{C^2}{n^2}+4Kt^2}, \qquad n=1,2,3,...
\label{eq:levels}
\end{equation}
In the continuum limit $n\to\infty$, the energy levels (\ref{eq:levels}) reduce to the familiar $1/n^2$ dependence of the Bohr formula. It is interesting to note that the same result is obtained by solving the continuum Hydrogen atom problem with a $C/r$ potential in the limit of infinitely many dimensions \cite{bohrmodel}.

The bound state energy levels of all other sectors can be obtained numerically as poles of the appropriate $G^F_k(\omega)$, given in Eq.~(\ref{eq:Gexact}).

\subsection{Scattering states}

A closer inspection of the Green's function (\ref{eq:Gexact}) reveals an imaginary part in the $|\omega|<2\sqrt{K}t$ range, corresponding to the continuum energy band, which follows the accumulation of bound states below $-2\sqrt{K}t$. Interestingly, the free particle problem \cite{Brinkman} has the same continuum spectrum, although its distribution of states within the band is different. It is thus instructive to first consider the case where $C=0$. For the time being, we restore the coordination number at the root to $z$, which makes the symmetric equal-amplitude superposition of all vertices an eigenstate with energy $-zt$. The lattice Green's function is given by the expression:
\begin{equation}
\mathcal{G}_\mathrm{free}(\omega)=\cfrac{1}{\omega-\cfrac{2zt^2}{\omega+\sqrt{\omega^2-4Kt^2}}}.
\label{eq:freeG}
\end{equation}
which indeed indicates a continuous spectrum between $-2\sqrt{K}t$ and $2\sqrt{K}t$. The gap from the $-zt$ uniform state to the lower edge of the band, $-2\sqrt{K}t$, can be explained by the peculiar dimensionality of the Bethe lattice. The imaginary part of Eq.~(\ref{eq:freeG}), which is proportional to the density of states $N(\omega)$, has a singularity at the edge, whereas the expected form is $N(\omega)\propto C(\omega-\epsilon)^{d/2}$ where $\epsilon$ is the edge of the band and $d$ is the dimension of the system. Given the Bethe lattice is ``infinitely dimensional'', $d\to\infty$, the band tails disappear, and the system acquires a gap. Introduction of closed loops in a way that results in a finite $d$ therefore leads to appearance of band tails stretching below $\epsilon=-2\sqrt{K}t$ towards the uniform state at $-zt$. 

\section{Summary and Discussion}
\label{sec:conclusion}

We have obtained an exact solution for the spectrum of the Coulomb potential problem on the $(K+1)$-coordinated Bethe lattice. The energy levels of different symmetry sectors are given by the poles of the \emph{forward hopping} Green's functions (\ref{eq:Gexact}). The sectors correspond to a choice of \emph{even} or one of the $(K-1)$ \emph{odd} quantum numbers for each generation of the Bethe lattice $n>1$. If the highest generation with an odd quantum number is $k$, then all of the states in the corresponding sectors have zero amplitude at nodes belonging to generations $n<k$, and the sectors' energy levels, obtained from the poles of $G_k^F(\omega)$, are $K^{k-2}$-fold degenerate. Each such sector can be mapped onto a hopping problem on an infinite half-line, where $V(n)$ is offset by $k$ and the hopping amplitude is multiplied by $\sqrt{K}$. This mapping exists for any problem whose Hamiltonian has the form (\ref{eq:Hgen}), and is particularly useful when, unlike in the Coulomb potential case, an exact solution cannot be obtained.

Models which involve approximating various physical structures as well as mathematical constructions by cycle-free graphs are plentiful. A single particle hopping on such graph can capture complex many body physics. For instance, the motion of a free particle on the Bethe lattice has been used to study antiferromagnets \cite{Brinkman} and water ice \cite{Chen}, whereas a particle hopping in the presence of random and radial potentials was used to model Anderson localization \cite{0022-3719-6-10-009} and spin ice \cite{diluted} respectively.

Apart from the few known exactly solvable cases, problems defined on Cayley trees are difficult to treat. While the exact diagonalizaion of a single particle problem may seem manageable using numerical methods, the large number of nodes at the boundary poses an immense difficulty. One may circumvent the issue by introducing closed cycles at the ends and/or focusing on the properties associated with the interior of the tree. We demonstrate that there is an alternative point of view for radial potentials. The mapping of each sector onto a one-dimensional problem allows for numerical treatment with tractable boundary effects stemming from the single edge node.

In addition to the broadly applicable mapping above, the exact solution derived in the present work may prove to be of use across various fields of physics, the inverse $r$ potentials being among the most common. In fact, the motivation behind this study came from considering the problem of emergent magnetic monopoles arising in a particular class of frustrated magnets -- spin ice -- where the Bethe lattice turns out to be a good approximation for the state space of the problem, and the Coulomb potential defined on it served to represent interactions between pairs of monopoles \cite{diluted}.

\section{Acknowledgments}

The authors thank S. L. Sondhi for useful discussions and collaboration on a related project \cite{diluted}, and Yen Ting Lin and Nikos Bagis for pointing us to Ramanujan's work on continued fractions. The authors acknowledge support of the Helmholtz Virtual Institute \emph{New States of Matter and their Excitations}, the Alexander von Humboldt Foundation (OP), and the German Science Foundation (DFG) via SFB 1143.

\bibliography{bethebib}

\begin{thebibliography}{12}
\expandafter\ifx\csname natexlab\endcsname\relax\def\natexlab#1{#1}\fi
\expandafter\ifx\csname bibnamefont\endcsname\relax
  \def\bibnamefont#1{#1}\fi
\expandafter\ifx\csname bibfnamefont\endcsname\relax
  \def\bibfnamefont#1{#1}\fi
\expandafter\ifx\csname citenamefont\endcsname\relax
  \def\citenamefont#1{#1}\fi
\expandafter\ifx\csname url\endcsname\relax
  \def\url#1{\texttt{#1}}\fi
\expandafter\ifx\csname urlprefix\endcsname\relax\def\urlprefix{URL }\fi
\providecommand{\bibinfo}[2]{#2}
\providecommand{\eprint}[2][]{\url{#2}}

\bibitem[{\citenamefont{Weiss}(1907)}]{Weiss1907}
\bibinfo{author}{\bibfnamefont{P.}~\bibnamefont{Weiss}}, \bibinfo{journal}{{J.
  Phys. Theor. Appl.}} \textbf{\bibinfo{volume}{6}}, \bibinfo{pages}{661}
  (\bibinfo{year}{1907}).

\bibitem[{\citenamefont{Bethe}(1935)}]{Bethe1935}
\bibinfo{author}{\bibfnamefont{H.~A.} \bibnamefont{Bethe}},
  \bibinfo{journal}{Proceedings of the Royal Society of London A: Mathematical,
  Physical and Engineering Sciences} \textbf{\bibinfo{volume}{150}},
  \bibinfo{pages}{552} (\bibinfo{year}{1935}), ISSN \bibinfo{issn}{0080-4630}.

\bibitem[{\citenamefont{Kurata et~al.}(1953)\citenamefont{Kurata, Kikuchi, and
  Watari}}]{Kurata1953}
\bibinfo{author}{\bibfnamefont{M.}~\bibnamefont{Kurata}},
  \bibinfo{author}{\bibfnamefont{R.}~\bibnamefont{Kikuchi}}, \bibnamefont{and}
  \bibinfo{author}{\bibfnamefont{T.}~\bibnamefont{Watari}},
  \bibinfo{journal}{The Journal of Chemical Physics}
  \textbf{\bibinfo{volume}{21}} (\bibinfo{year}{1953}).

\bibitem[{\citenamefont{Brinkman and Rice}(1970)}]{Brinkman}
\bibinfo{author}{\bibfnamefont{W.~F.} \bibnamefont{Brinkman}} \bibnamefont{and}
  \bibinfo{author}{\bibfnamefont{T.~M.} \bibnamefont{Rice}},
  \bibinfo{journal}{Phys. Rev. B} \textbf{\bibinfo{volume}{2}},
  \bibinfo{pages}{1324} (\bibinfo{year}{1970}).

\bibitem[{\citenamefont{Abou-Chacra et~al.}(1973)\citenamefont{Abou-Chacra,
  Thouless, and Anderson}}]{0022-3719-6-10-009}
\bibinfo{author}{\bibfnamefont{R.}~\bibnamefont{Abou-Chacra}},
  \bibinfo{author}{\bibfnamefont{D.~J.} \bibnamefont{Thouless}},
  \bibnamefont{and} \bibinfo{author}{\bibfnamefont{P.~W.}
  \bibnamefont{Anderson}}, \bibinfo{journal}{Journal of Physics C: Solid State
  Physics} \textbf{\bibinfo{volume}{6}}, \bibinfo{pages}{1734}
  (\bibinfo{year}{1973}).

\bibitem[{\citenamefont{Fisher and Essam}(1961)}]{percolation}
\bibinfo{author}{\bibfnamefont{M.~E.} \bibnamefont{Fisher}} \bibnamefont{and}
  \bibinfo{author}{\bibfnamefont{J.~W.} \bibnamefont{Essam}},
  \bibinfo{journal}{Journal of Mathematical Physics}
  \textbf{\bibinfo{volume}{2}} (\bibinfo{year}{1961}).

\bibitem[{\citenamefont{Chen et~al.}(1974)\citenamefont{Chen, Onsager, Bonner,
  and Nagle}}]{Chen}
\bibinfo{author}{\bibfnamefont{M.}~\bibnamefont{Chen}},
  \bibinfo{author}{\bibfnamefont{L.}~\bibnamefont{Onsager}},
  \bibinfo{author}{\bibfnamefont{J.}~\bibnamefont{Bonner}}, \bibnamefont{and}
  \bibinfo{author}{\bibfnamefont{J.}~\bibnamefont{Nagle}},
  \bibinfo{journal}{The Journal of Chemical Physics}
  \textbf{\bibinfo{volume}{60}} (\bibinfo{year}{1974}).

\bibitem[{\citenamefont{Gallinar}(1984)}]{Gallinar}
\bibinfo{author}{\bibfnamefont{J.-P.} \bibnamefont{Gallinar}},
  \bibinfo{journal}{Physics Letters A} \textbf{\bibinfo{volume}{103}},
  \bibinfo{pages}{72 } (\bibinfo{year}{1984}).

\bibitem[{\citenamefont{Petrova et~al.}(2015)\citenamefont{Petrova, Moessner,
  and Sondhi}}]{diluted}
\bibinfo{author}{\bibfnamefont{O.}~\bibnamefont{Petrova}},
  \bibinfo{author}{\bibfnamefont{R.}~\bibnamefont{Moessner}}, \bibnamefont{and}
  \bibinfo{author}{\bibfnamefont{S.~L.} \bibnamefont{Sondhi}},
  \bibinfo{journal}{Phys. Rev. B} \textbf{\bibinfo{volume}{92}},
  \bibinfo{pages}{100401} (\bibinfo{year}{2015}).

\bibitem[{\citenamefont{Aizenman et~al.}(2006)\citenamefont{Aizenman, Sims, and
  Warzel}}]{ASW}
\bibinfo{author}{\bibfnamefont{M.}~\bibnamefont{Aizenman}},
  \bibinfo{author}{\bibfnamefont{R.}~\bibnamefont{Sims}}, \bibnamefont{and}
  \bibinfo{author}{\bibfnamefont{S.}~\bibnamefont{Warzel}},
  \bibinfo{journal}{Contemporary Mathematics} \textbf{\bibinfo{volume}{412}},
  \bibinfo{pages}{1} (\bibinfo{year}{2006}).

\bibitem[{\citenamefont{Berndt}(1989)}]{Ramanujan}
\bibinfo{author}{\bibfnamefont{B.~C.} \bibnamefont{Berndt}},
  \emph{\bibinfo{title}{Ramanujan's Notebooks, Part II}}
  (\bibinfo{publisher}{Springer-Verlag}, \bibinfo{address}{New York},
  \bibinfo{year}{1989}), ISBN \bibinfo{isbn}{978-1-4612-4530-8},
  \bibinfo{note}{page 136}.

\bibitem[{\citenamefont{{Svidzinsky} et~al.}(2014)\citenamefont{{Svidzinsky},
  {Scully}, and {Herschbach}}}]{bohrmodel}
\bibinfo{author}{\bibfnamefont{A.}~\bibnamefont{{Svidzinsky}}},
  \bibinfo{author}{\bibfnamefont{M.}~\bibnamefont{{Scully}}}, \bibnamefont{and}
  \bibinfo{author}{\bibfnamefont{D.}~\bibnamefont{{Herschbach}}},
  \bibinfo{journal}{Physics Today} \textbf{\bibinfo{volume}{67}},
  \bibinfo{pages}{33} (\bibinfo{year}{2014}).

\end{thebibliography}

\end{document}